# Polymer nanoparticles to decrease thermal conductivity
## of phase change materials

Fabien Salaun

5    Laboratoire de Génie et Matériaux Textiles, UPRES EA 246, Ecole Nationale Supérieure des Arts et Industries Textiles, BP 30329, 59056 Roubaix Cedex 01, France

Pierre-Olivier Chapuis, Sourabh Kumar Saha, and Sebastian Volz[*]

10    Laboratoire d'Energétique Moléculaire et Macroscopique, Combustion, UPR CNRS 288, Ecole Centrale Paris, Grande Voie des Vignes, 92295 Châtenay Malabry, France

**Abstract**

15    Microparticles including paraffin are currently used for textiles coating in order to deaden thermal shocks. We will show that polymer nanoparticles embedded in those microsized capsules allow for decreasing the thermal conductivity of the coating and enhance the protection in the stationary regime. A reasonable volume fraction of polymer nanoparticles reduces the conductivity more than predicted by Maxwell

20    mixing rules. Besides, measurements prove that the polymer nanoparticles do not affect the latent heat and even improve the phase change behaviour as well as the mechanical properties.

25    Keywords: polymer nanoparticles, phase change microparticles, microencapsulation

---

[*] Tél :33 14113 1049, Fax : 33 1 4702 8035, volz@em2c.ecp.fr

## 1. Introduction

In the two past decades, microencapsulated Phase Change Materials (PCM) have drawn an increasing interest to provide enhanced thermal functionalities in a wide variety of applications [1-4]. When the encapsulated PCM is heated to the melting point, it absorbs heat as it goes from a solid state to a liquid state. This phase change slows down the temperature increase. It can be applied to clothes technology, building insulation, energy storage as well as to coolant liquids. On a more general basis, it can be used to design a broad variety of thermal transient regimes.

The PCMs used for ambient temperature related applications are carbohydrates with different chain lengths [5-8] or paraffins [9-10]. The PCMs are encapsulated in small spheres in order to be contained in a liquid state. The microcapsules possess approximate diameter of 1-10 μm and are resistant to abrasion, pressure, heat and chemicals according to the shell chemical compounds [11-12] Micro-sized capsules are required to provide a large contact area with the environment and ensure an optimal efficiency of the phase change.

Among a multitude of possible wall materials for microcapsules, amino resins, more especially melamine-formaldehyde, play a main role, more precisely in the patent literature [24]. Amino resins represent an interesting economical alternative, since these polymer raw materials have been produced on a large scale for many years, and since they have already been practised in many processes like phase separation and interfacial reaction (e.g. dicarboxylic acid dichlorides and di or triamines). Furthermore melamine-formaldehyde microcapsules prepared by in situ polymerisation have impermeable shell.

Up to now, most studies have targeted microcapsules, but encapsulation of micro-nanospheres remains an innovative field [25]. In this work, we show how polymer nanoparticles (NP) embedded in PCM microcapsules significantly decreases the thermal conductivity to improve the thermal barrier effect. We also prove that the modified microcapsules have better mechanical properties and even an enhanced phase change behavior.

60   The fabrication of the simple core-shell PCM and of the polymer NPs based
microcapsules are presented in the second section. The thermal measurements are
described in the third paragraph. The phase change behaviour was characterized by
conventional Differential Scanning Calorimetry to measure the latent heat and the
temperature range of the solid-liquid transition. Both materials exhibited the same
65   latent heat but the phase change occurs on a wider temperature interval in the case of
the NP based PCM. The thermal conductivity was measured with a Scanning
Thermal Microscope because the conventional methods are difficult to apply.
Contact technique such as hot guarded plates introduce thermal contact resistances
which are difficult to estimate and the low optical absorption of porous materials
70   makes the application of optical techniques intricate. The value of the thermal
conductivity of the polymer NPs was found to be extremely low, i.e. only twice the
one of air. The addition of a reasonable fraction of NPs in the PCM capsule results in
a decrease by a factor of three of the effective thermal conductivity. The outcome is
a significant improvement of the thermal barrier in the steady state. The
75   mechanical properties were estimated on a qualitative basis by using an atomic force
microscope. The damage threshold was reached in the basic microcapsule before the
one of the modified materials. The results are described and commented in the last
part.

80   **2. Preparation of microcapsules**

The choice of the process of synthesis is generally dictated by the chemical nature of
the polymer forming the membrane and by the desired mean diameter. The thermal
properties of a microcapsule are influenced by three factors: its mean diameter, its
85   expansion during the phase change process and its shell's chemical nature. In his
study, Bryant noticed that the smaller the mean diameter was, the more the weight
loss was significant at high temperature [16]. The microcapsule wall is greatly
affected by the phase change process, since between the melting and the
crystallization, the volume varies with an approximate magnitude of 10 %. This can
90   induce the diffusion of the core material at high temperature. Most of synthesized
microcapsules are intended for application fields at a temperature lower than 150 °C.
Therefore, microcapsules containing phase change materials have yet been
synthesized with urea-formaldehyde [17], urea-formaldehyde-resorcinol [18],

polyamide [19], gelatin-formaldehyde [20], polyurethane [21], melamine-
95 formaldehyde [22] and urea-melamine-formaldehyde [23] shell. In this last study, the
factors improving thermal stability were the addition of cyclohexane to the n-
octadecane and the copolymerization of urea with melamine and formaldehyde.

The paraffins N-hexadecane, n-eicosane, tetraethyl orthosilicate and sodium
100 phosphate dodecahydrate used as PCMs formulation were purchased from Acros
organics. Melamine-formaldehyde resin (Arkofix® NM) used as shell-forming was
kindly supplied by Clariant (France). Poly(vinyl alcohol) (95 % hydrolysed, average
M.W. 95,000) and methylene diisocyanate (Suprasec 2030, Huntsman ICI) were
employed to entrap inorganic salt. Nonionic surfactants, Tween® 20 (Acros
105 Organics, France), Span® 85 and Polyethylene glycol 400 dioleate (Aldrich, France)
were used as emulsifiers. For pH control, triéthanolamine and citric acid were used
(Aldrich, France).

Our technique of microencapsulation for the simple core shell structure consists in a
three step process. The first step is the crucial phase of emulsion. All physical
110 properties of the microcapsules will be established at that time (e.g. size distribution,
morphology, shape, …) [13-14]. Indeed, the dispersion of the organic phase in the
continuous phase is prepared by a fragmentation process where wide areas of oil are
broken up by shear or elongation. Therefore the droplet sizes are highly affected by
the physicochemical properties of the two immiscible liquids (e.g. viscosities,
115 densities, pH, interfacial tension, temperature, …) and by the agitation system
(especially stir speed and shape of stirrer) [15]. The preparation of our microparticles
was carried out according to the method of *in situ* polymerisation of melamine-
formol. The inner phase was emulsified in an aqueous solution containing 4 g of
Tween® 20 in 100 g of water and 9.2 g amino resin.

120 In the second phase, the pH was reduced to 4 by adding acid citric solution, at a
stirring rate of 8000 rpm, at room temperature, during 3 minutes with a homogenizer.
The PCM core particles are generated in this phase.

Finally, the reaction mixture was heated at 60°C, stirring was continued using a
blade stirrer at low speed (400 rpm, RW20, IKA, Germany) for 4 hours until the end
125 of the polycondensation. The microcapsules were recovered by filtration, washed
with methanol and water, and dried at room temperature during a night.

Figure 1 reports a Scanning Electron Microscope (SEM) image of the resulting core-shell microcapsules named 'ARD'. The size distribution appears as quite homogenous and the particle diameter ranges from 2 to 4 micrometers. The composition of each inner phase used in the preparation is listed in Table 1.

130

To produce polymer nanoparticles in the basic core shell structure, the first step consists in generating polymer spheres in paraffin. In this aim, we mixed 4 g of sodium phosphate dodecahydrate and 2 g of water. The solution is added to 0.5 g of mixture of nonionic sufactants in 7 g of n-alkane. After stirring during 15 min, the droplet size was reduced by homogenizing the emulsion during 15 min at 9500 rpm with an Ultra-Turrax homogenizer (Ika, Germany). In the same way, another emulsion was prepared by homogenizing 8 g of 95 % hydrolysed poly(vinyl alcohol) solution (5 wt.-%) in 8 g of n-alkane. The polymer spheres were prepared by shearing under high speed the two emulsions with 3 g of MDI to crosslink the shell at 50°C for 30 min. The resulting polymer nanoparticles were observed under SEM as illustrated in Figures 2. The mean diameter is about 50 nanometers and the scattering in size seems neglectible. The porous material formed by those particles is labeled '18' in the following.

135

140

In the second and last step, the previous mixture is introduced in the place of the basic PCM in the microencapsulation process described above. The surface morphology of the final product was observed with an optical microscope (Zeiss Axisokop) in the phase difference mode as shown in Figure 3 (top). The inner structure of the microcapsule is revealed in Figures 3 (bottom). A rather compact distribution of polymer nanoparticles is shown in both images.

145

150

## 3. Results and discussion

### 3.1. Latent heat

155 A conventional Different Scanning Calorimetry (DSC) was performed to prove that the polymer nanoparticles do not affect the latent heat. The thermal behavior of the microcapsules was recorded to analyse the influence of nanoparticles on latent heat storage, using a TA instrument type DSC 2920 piloted on PC with TA Advantage control software. The measurements were performed by using increase rates for the temperature ranging from 0.5 °C/min to 20 °C/min. The values of latent heat were

160

found equal within the measurement accuracy for both pure paraffin microparticles and polymer nanoparticles based microcapsules. The value of the latent heat is equal to 176 J/g. The DSC data reported in Figure 4 however indicate a difference in the phase change behaviour. The temperature at which the solid-liquid transition starts is

165   the same but the temperature interval is 50% larger for the NPs based PCM compared to the conventional mirocapsules. The phase change will hence be effective on a larger temperature range. We presume that interaction between nanoparticle surface and paraffin induces a specific molecular structure in paraffin. An extra energy is necessary to break this molecular arrangement. This modification

170   in the phase change is of course beneficial for textile application because temperature will be maintained under a larger heat flux amplitude.

## 3.2. Thermal conductivity

The thermal conductivity is estimated by using a scanning thermal microscope. It

175   consists in a conventional atomic force microscope mounted with a hot wire probe as illustrated in Figure 5. The probe is a wollaston wire made of a platinum core 5 microns in diameter and a silver coating 70 microns in diameter. The silver coating is etched to uncover the platinum wire over a length of 2L=200 microns. This tip was studied in several of our previous works [8]. A modulated (AC) electrical current is

180   used to Joule heat the wire and the second harmonic of the temperature is measured [9].

The wire temperature is related to the heat flux flowing from the tip to the sample. And this heat flux between the tip and the sample is driven by a global thermal conductance $G_{eq}$ including the contributions of the contact conductance and the one

185   of the sample $G_s$.

We perform a two-step approach to identify the ratio between two thermal conductivities $\lambda_{S1}$ and $\lambda_{S2}$ of two different samples. Measurements are performed in ambient and in vacuum when varying contact force F between sample and tip. Those measurements provide the force derivatives of the sample conductance and the one

190   of $G_{eq}$. We assume that the contact conductance and the contact hardness do not vary significantly when probing the two dissimilar samples. In those conditions, it can be shown that the ratio between both thermal conductivities arises as follows:

$$\frac{\lambda_{S1}}{\lambda_{S2}} = \left( \frac{dG_S}{dF}\bigg|_1 \middle/ \frac{dG_S}{dF}\bigg|_2 \right)^{1/2} \left( \frac{dG_{eq}}{dF}\bigg|_2 \middle/ \frac{dG_{eq}}{dF}\bigg|_1 \right)^{1/4} \tag{1}$$

Note that Eq. (1) does not rely on the determination of a specific mechanical or thermodynamic property.

Figure 6 reports on the thermal conductance $G_{eq}$ against force for samples 18 and ARD. The slopes are not clearly different showing that the contact conductance is predominant over the sample conductances. But the difference in levels indicate that the paraffin in sample ARD is significantly more conductive than sample 18 including polymer. Eq. (1) provides the ratio of $\lambda_{18}/\lambda_{ARD}=0.31$. From the reference data $\lambda_{ARD}=0.26$ W.m$^{-1}$.K$^{-1}$ and the very low thermal conductivity of $\lambda_{18}=0.08$ W.m$^{-1}$.K$^{-1}$ is found to be only three times the one of air.

Figure 7 presents the evolution of the thermal conductance $G_{eq}$ as a function of the electrical power in the probe. The hotter the probe is, the more resistive samples are. The linear behavior suggests that the samples might melt and become more resistive. The relative values of the samples thermal conductivities are reported in the insert of Figure 7. Microparticle based samples, that are referred as E2 H/E and 18, are 60% to 70% more insulating than paraffin. This confirms the impact of the very insulating polymer nanoparticles on the PCM thermal conductivity.

Samples E2 and H/E include 68% of paraffin and 32% of polymer. Consequently, Maxwell-Garnett mixing rules predict a thermal conductivity decrease of 23%, which is about threefold less the experimental data of 60%. The thermal resistances at interfaces between polymer nanoparticles and paraffin might explain such a discrepancy. A volume fraction of 10% also corresponds to an average inter-particle distance of 0.35 time the particle diameter, i.e. about 15nm. We therefore suggest that percolating networks of polymer nanoparticles might be responsible for the drastic reduction in heat conduction.

This property implies a considerable gain, especially in the field of textile because the temperature drop between the inner and the outer face is increased threefold.

## 3.3. Bulk modulus

The Atomic Force Microscope allows for probing the mechanical properties of the surface. Figure 8, provides the force applied on the tip versus the cantilever altitude.

225     The stiffer the material is, the larger the slope is. Sample 18 is the stiffest material because it does not include any paraffin. The E2 sample has a lower melting point than H/E. The corresponding slope is the same as the one of sample 18 because we presume that the polymer contribution predominates because of its structural network included in a solid matrix. In sample H/E, paraffin is in the liquid phase at ambient

230     which might explain the slight decrease of the slope. Isolated polymer nanoparticles might indeed slide among the aggregates. The ARD sample has the same slope as the H/E one in the range of the small forces because they include the same paraffin. The data point corresponding to the largest force seems to indicate that the capsule shell was broken. This mechanism appears only in the case of sample ARD because the

235     polymer structure does not enforce the capsule.

## 4. Conclusion

    We have fabricated core shell PCMs including polymer nanoparticles and have characterized their structural configuration via Scanning Electron Microscopy. The

240     diameter of polymer nanoparticles is in the vicinity of 50nms and the shell size is of a few microns. We have shown that the addition of polymer nanoparticles has a very beneficial impact on thermal properties. Firstly, we have proven by using usual DSC that the temperature interval of the phase change is augmented by 50%. This significantly improves the role of thermal barrier of the PCM. We suppose that this

245     effect is related to the ordered structure of paraffin molecules around the polymer nanoparticles. This configuration might require a larger amount of energy to start for melting.

    Secondly, the thermal conductivity of polymer based PCM is decreased by 60% in comparison with the pure paraffin sample. This reduction is threefold larger than the

250     one predicted by conventional Maxwell-Garnett mixing rules. We have highlighted that the average separation distance between particles is about 15nm only and aggregation occurs as proven by the SEM snapshots. The network of polymer nanoparticles might break the percolation of paraffin and hence explains this very special behavior.

255

320

Captions

Figure 1: Scanning Electron Microscope image of the ARD sample. The spherical shell structure has a size of 1 to 3 microns.

325

Figure 2: Scanning Electron Microscope image of sample 18. The polymer nanoparticles have diameters of about 50nms. The high particle density allows for aggregation.

330 Figure 3: Optical (top) and Scanning Electron Microscope (bottom) images of sample E2 proving the high density and the configuration of polymer nanoparticles inside the shell. Although shells include 68% of paraffin, the nanosized polymer particles generates a very dense and thermally insulating structure.

335 Figure 4: DSC characterization of the E2 sample showing the extension in the temperature interval of the phase change. Several increase rates for the temperature are displayed. Each of them leads to the same latent heat.

Figure 5: Sketch of the thermal probe. A wollaston wire is formed as a tip. A mirror
340 is stuck on the top face to allow for the detection of the tip deflection. The electrical resistance of the wire is measured to deduce its temperature and its dissipated heat flux.

Figure 6a,b: (top) Thermal conductance $G_{eq}$ related to the heat flux exchanged
345 between the thermal probe and the sample versus contact force between tip and sample. (bottom) Thermal conductance $G_s$ of the sample versus contact force between tip and sample. The force is proportional to the tip deflection, which is measured by the photodiode presented in Figure 5. The force units are hence provided by the photodiode signal in NanoAmperes.

350

Figure 7: Thermal conductance $G_{eq}$ related to the heat flux exchanged between the thermal probe and the sample versus the input electrical tip voltage squared. The insert represents the normalized thermal conductivities of the four samples.

355    Figure 8: Contact force between thermal probe and sample as a function of the tip
       height. The tip height is measured by the piezocrystal voltage.

**Table 1 Typical composition of inner phase for preparation of microparticles**

| Nomenclature | Compound wt.-% |
|---|---|
| ARD | n-eicosane / n-hexadecane / tetraethyl orthosilicate<br>48 / 48 / 4 |
| 18 | PVA-MDI microsphères*<br>100 |
| E2 | n-eicosane / PVA-MDI microsphères*<br>68 / 32 |
| H/E | n-eicosane / n-hexadecane / PVA-MDI microspheres*<br>34 / 34 /32 |

360

365

Figure 1

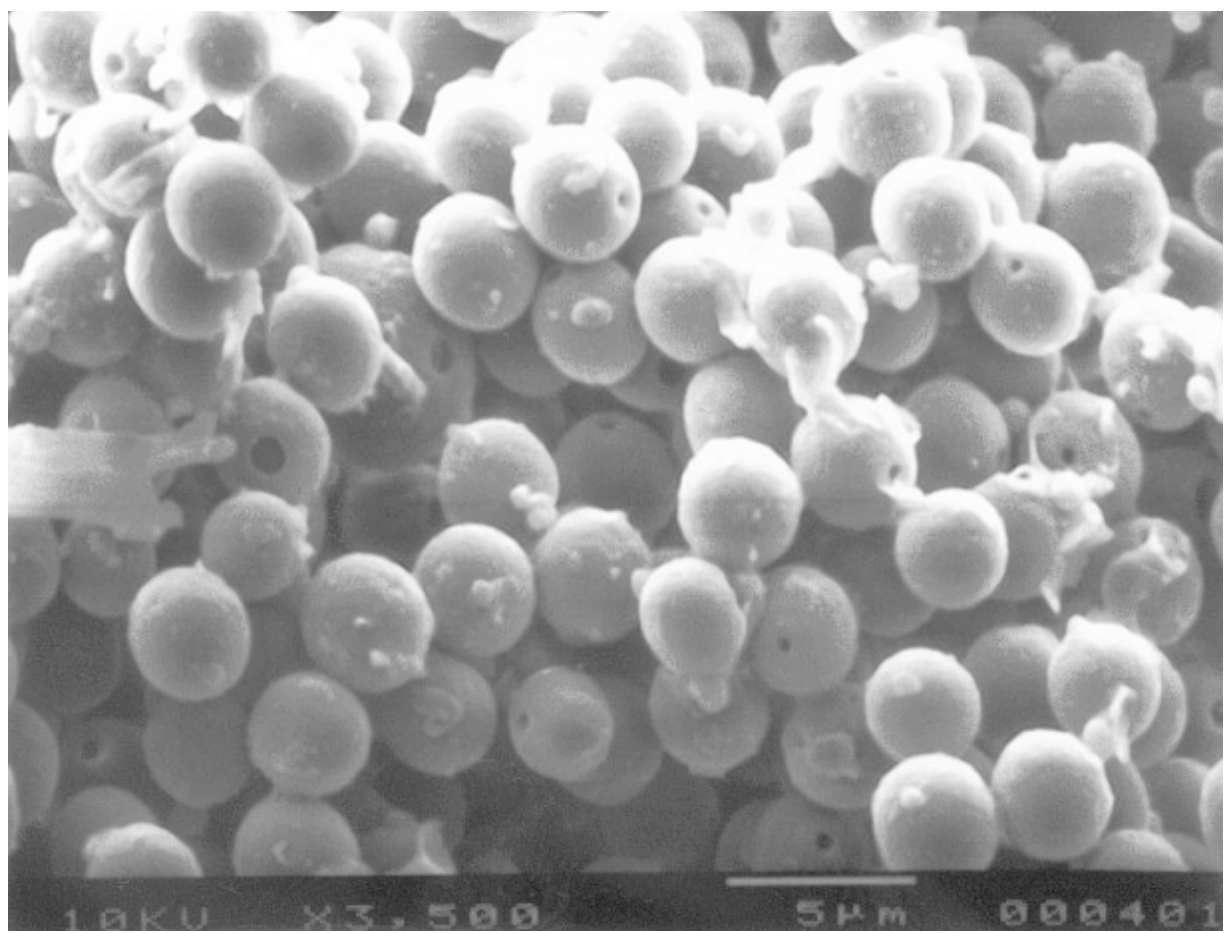

370

Figure 2

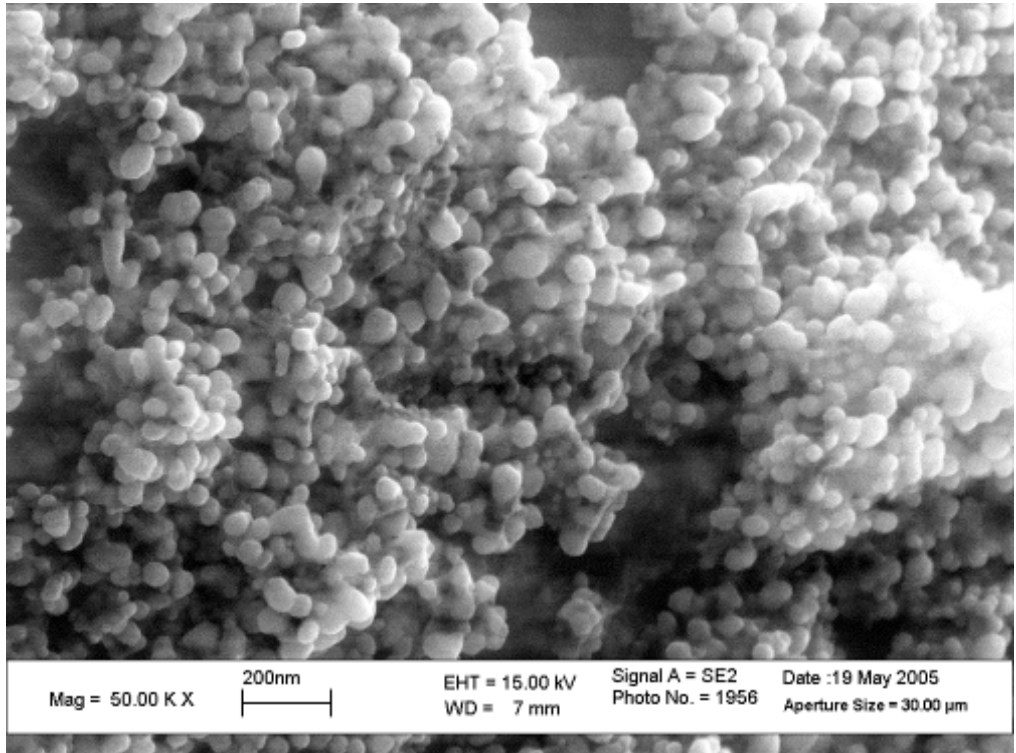

375

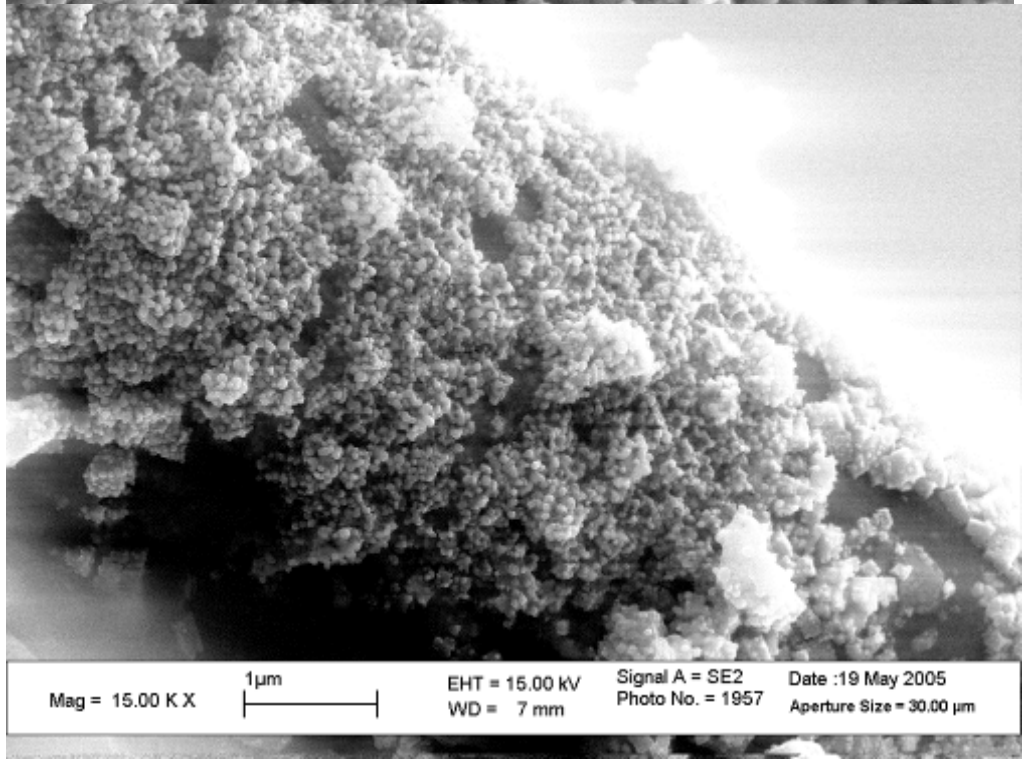

380    Figure 3

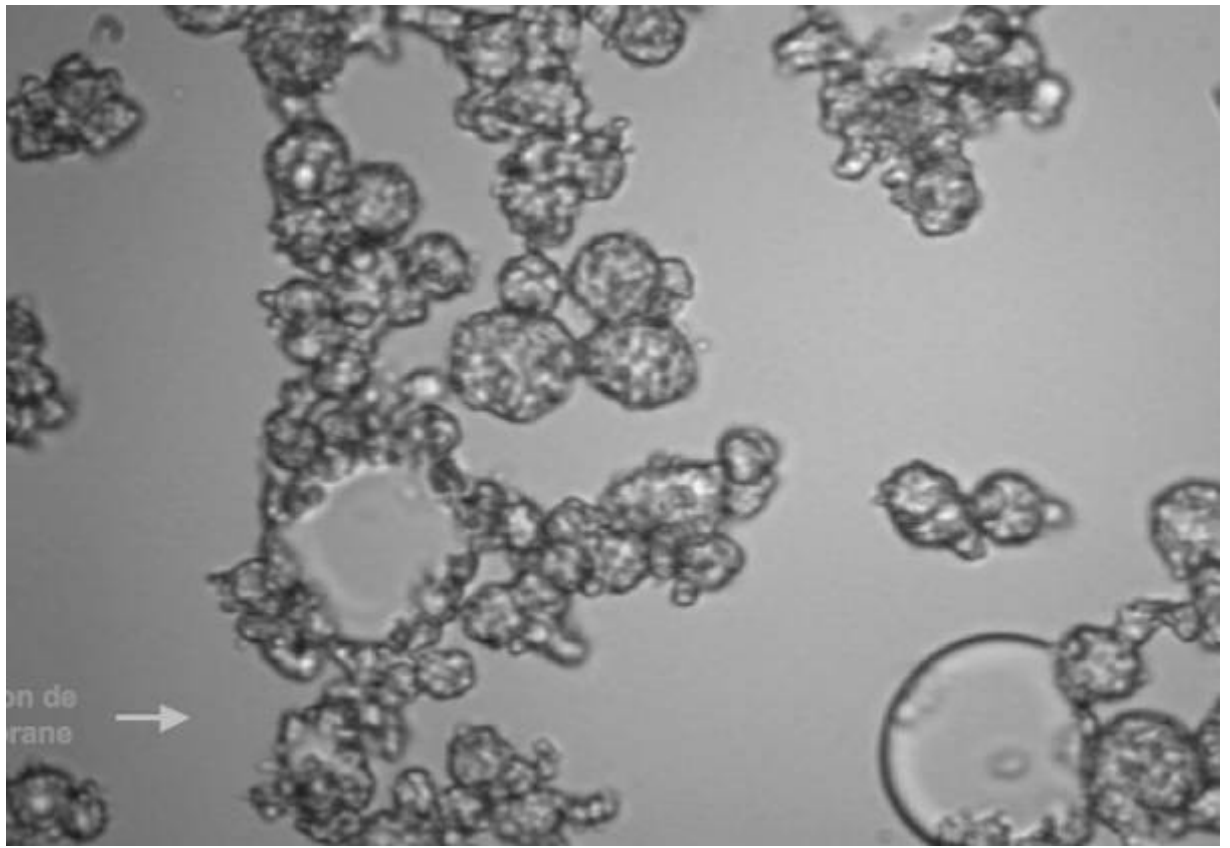

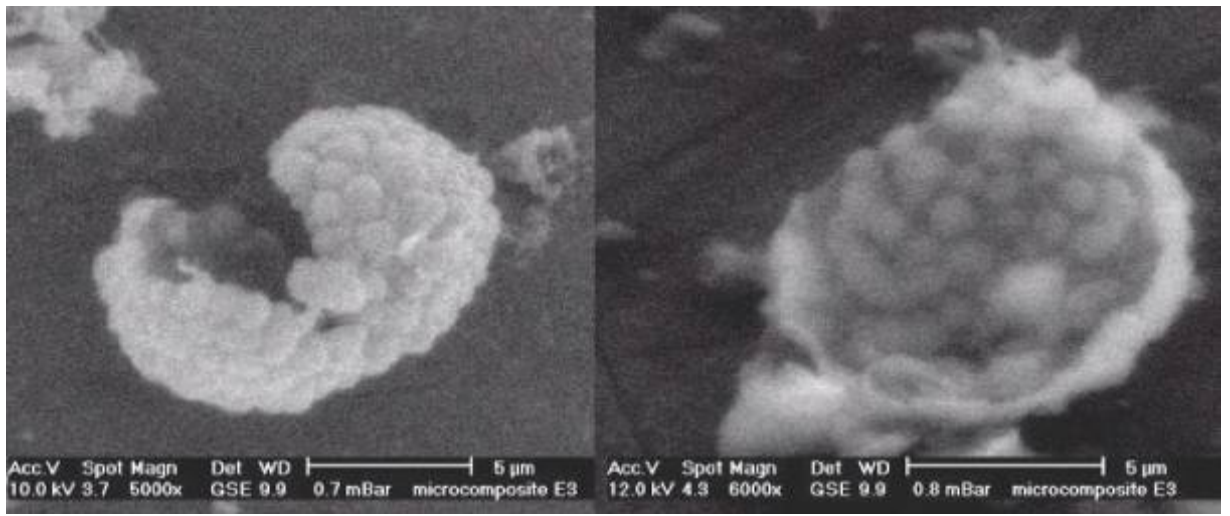

385

Figure 4

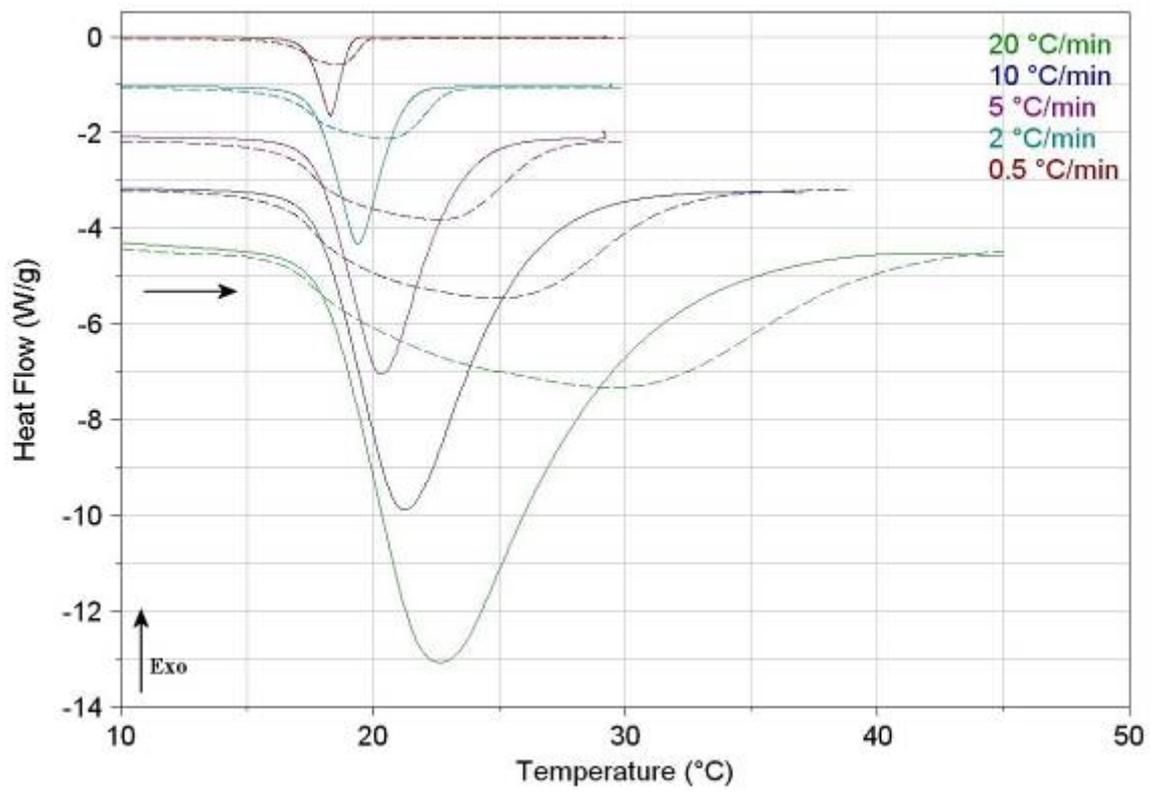

390    Figure 5

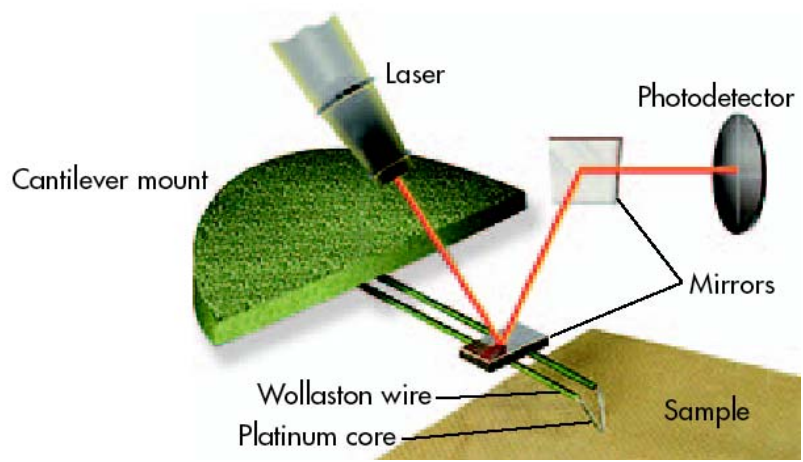

395

Figure 6

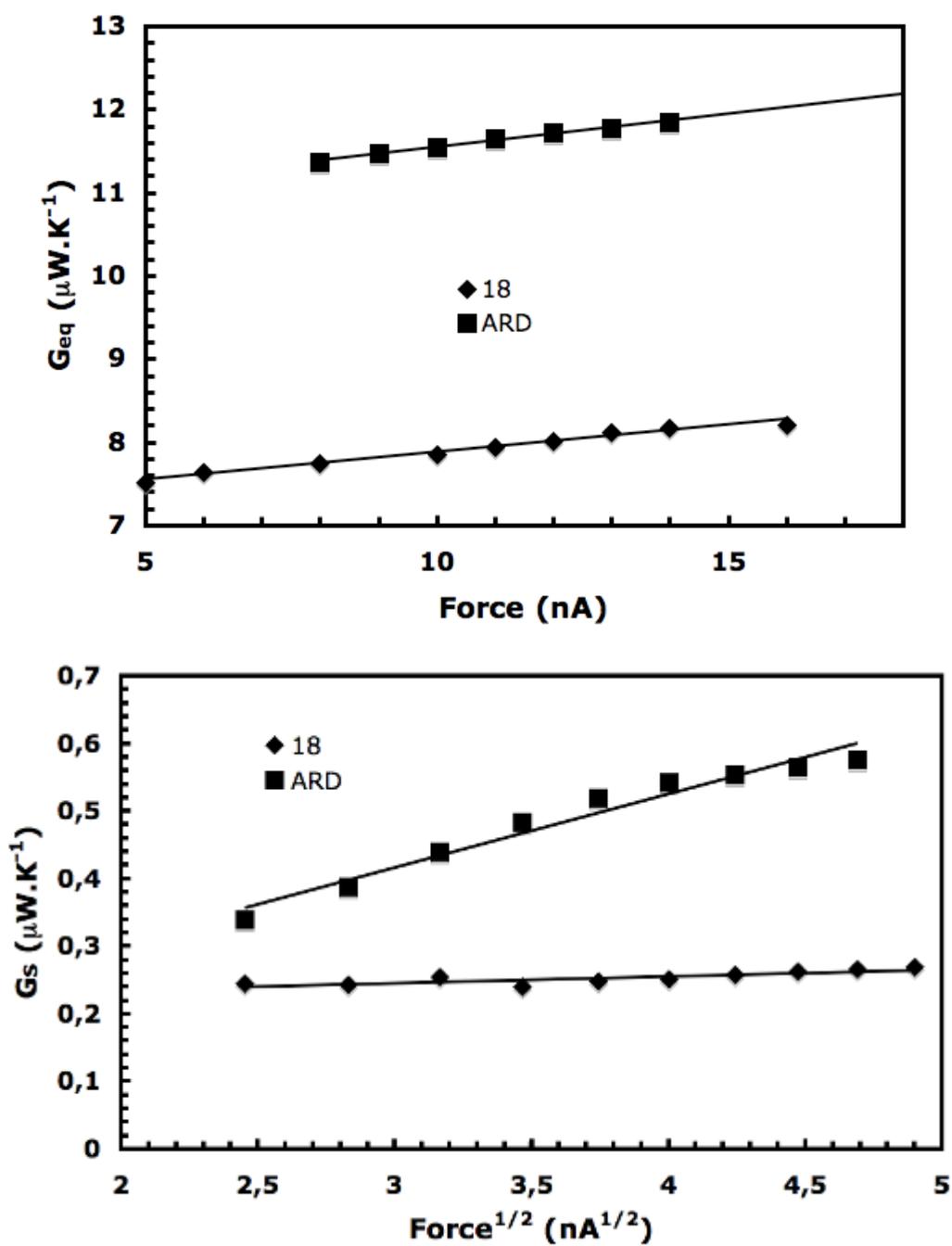

Figure 7

405

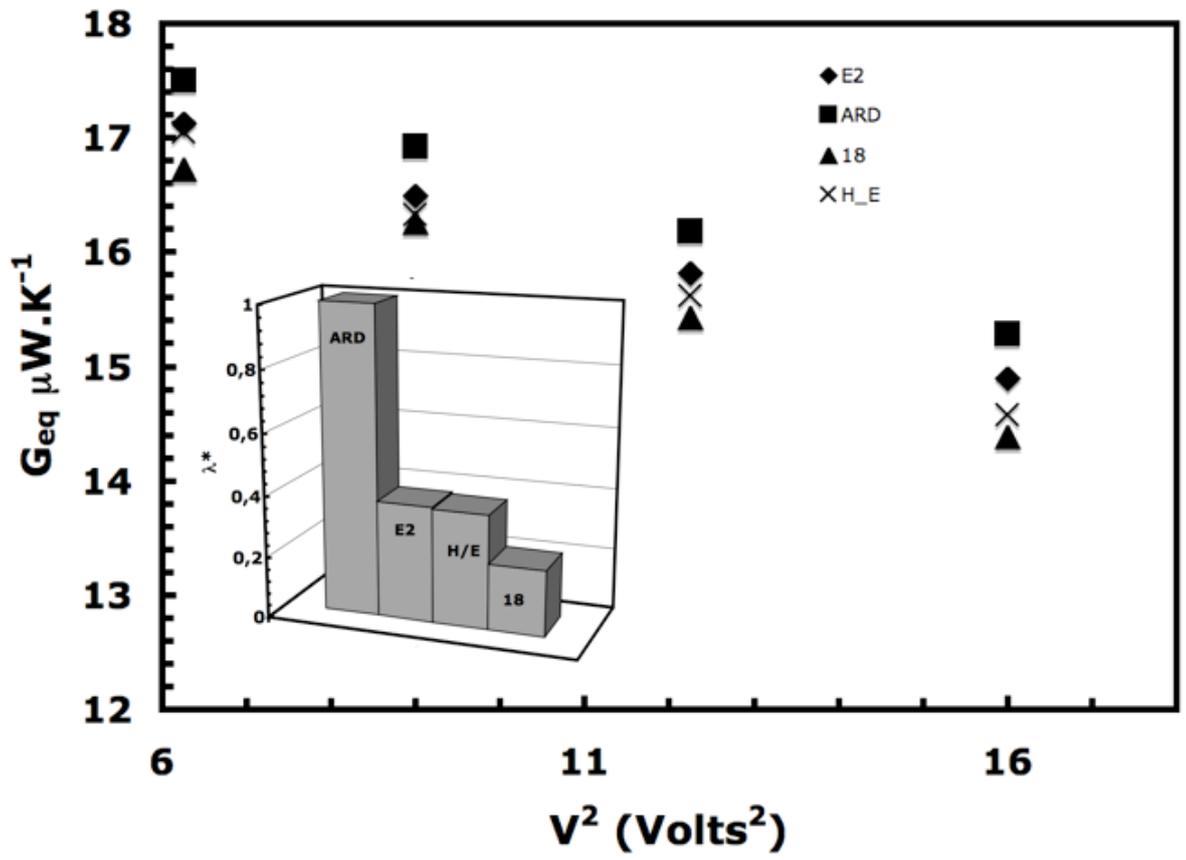

410

Figure 8

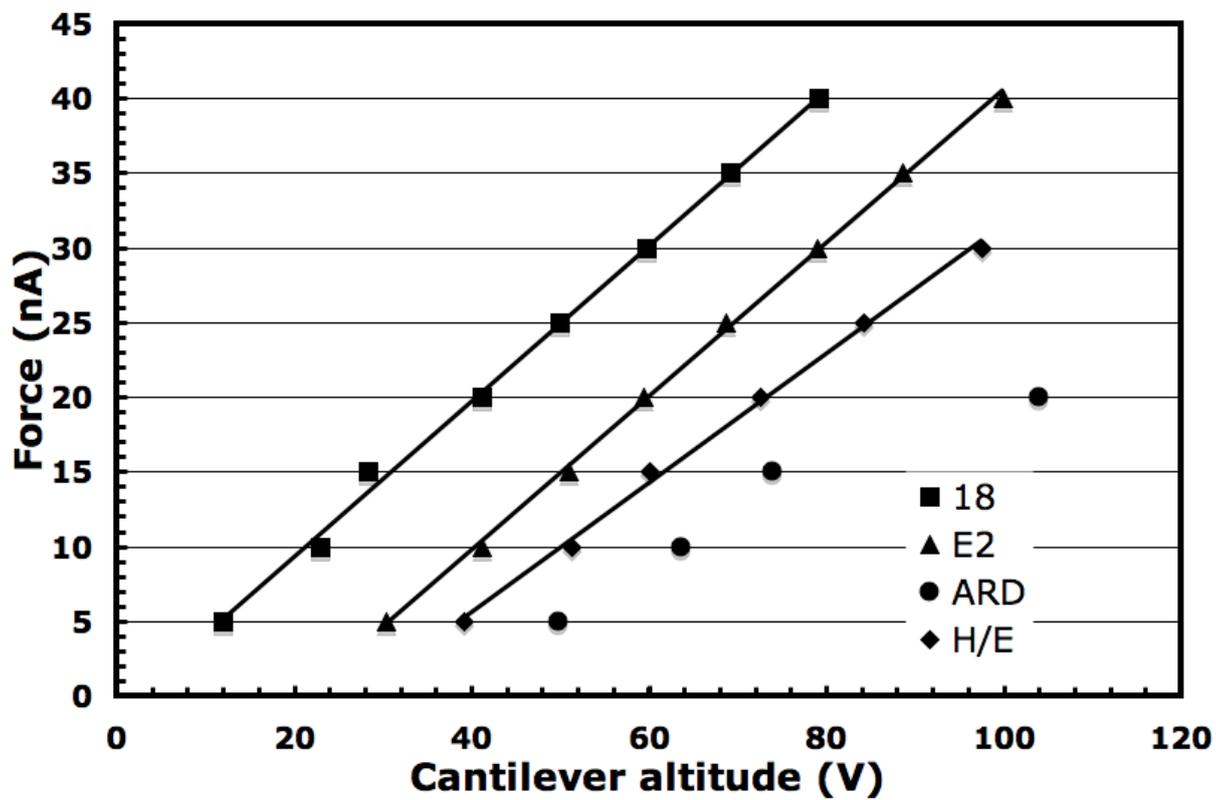